# Weak-field FMR and magnetization near the collinear to conical ferrimagnetic phase transition in the U-type hexaferrite $Sr_4CoZnFe_{36}O_{60}$ ceramics

M. Kempa,[1a] V. Bovtun,[1] D. Repček,[1,2] J. Buršík,[3] S. Kamba[1]

[1]Institute of Physics of the Czech Academy of Sciences, Na Slovance 2, 182 00 Prague, Czech Republic

[2]Faculty of Nuclear Sciences and Physical Engineering, Czech Technical University in Prague, Břehová 7, 115 19 Prague, Czech Republic

[3]Institute of Inorganic Chemistry, Czech Academy of Sciences, 250 68 Řež near Prague, Czech Republic

[a]Corresponding author: kempa@fzu.cz

**Abstract.**

Temperature evolution of the ferromagnetic resonance (FMR) and its interference with other microwave (MW) magnons near the collinear to conical ferrimagnetic phase transition at $T_{c2}$ = 305 K is found to correlate with the evolution of the weak-field magnetization: from two components in the conical phase to one component in the collinear phase. The FMR splitting above $T_{c2}$ correlates with the splitting of the coercive fields of the two magnetization components. A high sensitivity of the FMR to the weak magnetic bias near $T_{c2}$ is shown to be caused by the gradual transformation of the conical spin magnetic moments to the longitudinal ones. Application of the weak magnetic bias allows to adjust the MW absorption, and its level of above 30 dB is achieved near the FMR frequency (5.7 – 7.2 GHz), that allows to consider the $Sr_4CoZnFe_{36}O_{60}$ hexaferrite ceramics as a possible MW absorbing material.

## 1. Introduction

Hexaferrites exhibit a set of magnetic structures (collinear, conical) depending on the crystal structure, chemical composition, temperature and applied magnetic field. [1,2]. Due to their magnetic properties, hexaferrites find a variety of the practical applications in electrotechnics and electronics, including the microwave (MW) electronics [3,4]. Recently, we reported on our studies of the magnetization, high-frequency magnetic and dielectric properties [5], as well as of the MW magnetic excitations [6] of the new U-type hexaferrite $Sr_4CoZnFe_{36}O_{60}$ ceramics (CoZnU). This material was found to exhibit three magnetic phase transitions on cooling [5]: from a paramagnetic to a collinear ferrimagnetic structure at $T_{c1}$ = 635 K, then into a first conical magnetic structure (probably longitudinal) at $T_{c2}$ = 305 K, and at $T_{c3}$ = 145 K into a second conical magnetic structure (probably transverse). The shape of the hysteresis loops in the conical phases was shown to differ from those in the collinear phase. The phase transitions at $T_{c2}$ and $T_{c3}$ are characterized by the essential anomalies of the magnetic permeability, while no anomalies of the dielectric permittivity are observed. Magnetoelectric and polarization measurements revealed no magnetoelectric effect and ferroelectric polarization [5], so CoZnU is surprisingly not multiferroic, unlike the related $Sr_4Co_2Fe_{36}O_{60}$ [7]. The 9 MW excitations were revealed between 100 MHz and 35 GHz and assigned to dynamics of the magnetic domain walls and inhomogeneous magnetic structure of the ceramics, to the natural ferromagnetic resonance (FMR) and other magnons, and to the magnetodielectric modes. Using experimental data from [5,6], we analyse below in detail the temperature dependence of the FMR near $T_{c2}$ without application of the bias magnetic field and its change under application of the weak bias ($H$ < 700 Oe) at room temperature (i.e., in the conical phase close to $T_{c2}$). Correlation of the FMR behaviour with the weak-field magnetization is discussed, as well as the interference of the FMR with the nearest MW modes. Details of the CoZnU preparation, structure and experimental techniques used for the characterization are described in [5,6].

## 2. Temperature evolution of the weak-field magnetization

In spite of the gradual evolution of the magnetization $M(H)$ loop shape with increasing temperature through the conical to collinear magnetic phase transition at $T_{c2}$, the common feature remains in both phases: a strong increase of the magnetization at low applied magnetic field ($H$ < 300 Oe), see Fig 1a and [5]. It evidences the high value of the weak-field magnetization $M_W$. Presence of two $M_W$ contributions, $M_{W1}$ and $M_{W2}$, was revealed [5]. Temperature dependence of the weak-field $M_{W1}$ can be fitted in the collinear phase to a phenomenological power law [8]:

$$M(T) = M_0 \left(1 - \frac{T}{T_0}\right)^\beta, \qquad (1)$$

with $T_0 = 640$ K, $\beta = 0.25$ and $M_0 = 5.2$ $\mu_B$/f.u. (see a red dashed line in Fig. 2a, detailed fit of $M_{W1}$ up to 640 K was published in [5]). The $M_{W1}(T)$ can explain only a half of the $M_W$ value in the conical phase, therefore the second $M_{W2}$ contribution is necessary.

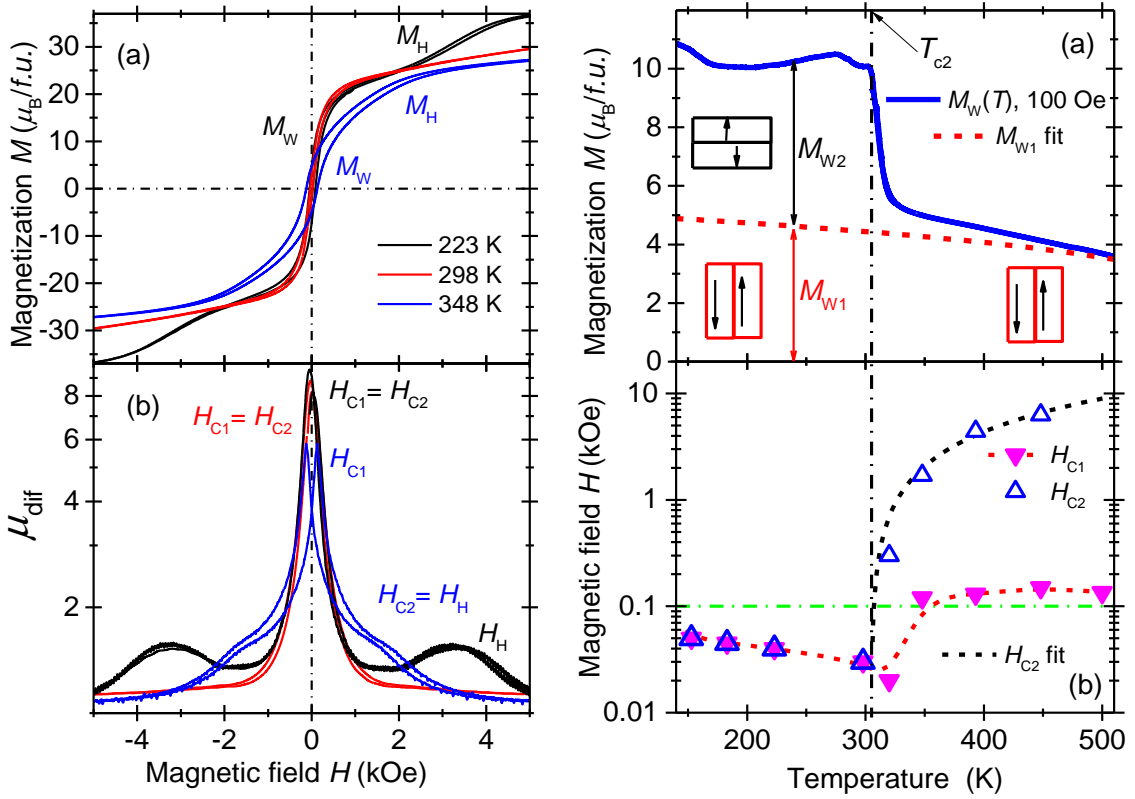

Fig. 1. Typical (a) magnetization hysteresis loops showing the weak- and high-field magnetization contributions ($M_W$ and $M_H$) in the conical and collinear ferrimagnetic phases; and (b) differential permeability hysteresis loops with two maxima, corresponding to their coercive fields ($H_{C1}$, $H_{C2}$ and $H_H$).

Fig. 2. Temperature dependences (a) of the magnetization in the weak magnetic field $M_W(T)$ consisting of two contributions and (b) of their saturation coercive fields $H_{C1}$ and $H_{C2}$. The black dashed line corresponds to the $H_{C2}$ fit (Eq. 3). Objects in (a) schematically illustrate the domain structures in the conical and collinear ferrimagnetic phases. Black dash-dot vertical line corresponds to $T_{c2}$. A horizontal green dash-dot line in (b) corresponds to 100 Oe, i.e. to the magnetic field of the $M_W(T)$ in (a).

The saturation fields $H_S$ of the weak- and high-field magnetizations $M_W$ and $M_H$ were estimated in [5]. This estimation was not obvious, especially in the collinear phase. Therefore, we determine the coercive fields using the differential static magnetic permeability $\mu_{dif}$ (Fig. 1b), obtained from the magnetization loops [5] as

$$\mu_{dif} = 1 + \frac{dM}{dH}, \qquad (2)$$

where $M$ - magnetization in A/m, $H$ – applied magnetic field in A/m. The $\mu_{dif}(H)$ maximum corresponds to the maximum slope of the $M(H)$ curve which is observed at the initial coercive field $H_C$ where $M = 0$ [9]. In this way, the magnetization contributions can be distinguished, and their coercive fields can be reliably determined (see $H_{C1}$, $H_{C2}$ and $H_H$ maxima in Fig. 1b corresponding to the weak- and high-field magnetizations ($M_W$ and $M_H$) in Fig. 1a).

In the conical phase, both contributions to the weak-field magnetization are characterized by the low coercive ($H_{C1} \approx H_{C2} < 50$ Oe) fields. In the collinear phase, $H_{C1}$ and $H_{C2}$ split. $H_{C1}$ remains at the low level, $H_{C2}$ rapidly increases on heating above 320 K (Fig. 2b). It explains a gradual extinction of $M_{W2}$ in the collinear phase (Fig.2a). The main $M_{W2}$ decrease takes place between $T_{c2}$ and 320 K. Above 320 K, $H_{C2}$ should be mainly related to the high-field magnetization. $H_{C1}(T)$ has a minimum near $T_{c2}$ and then increases above 100 Oe in the collinear phase.

The saturation value of magnetization achieved at high magnetic fields (~10 kOe) is ~5 times higher than the $M_W$ value at all temperatures below 580 K, i.e. in all ferrimagnetic phases [5]. Therefore, we assume that $M_W$ originates from the partial rearrangement of the complicated domain structure and from the partial re-ordering (re-orientation) of the spin moments in the ceramics with randomly oriented grains. This assumption explains why $M_W$ is roughly the same in both conical phases: transverse and longitudinal conical domains contribute to $M_W$ due to the random orientation of grains. In the collinear phase, mainly rearrangement of the collinear domains with the low coercive field (i.e., $M_{W1}$) contributes to $M_W$. Schematically, the change of the domain structure is shown in Fig.2a. Presence of the $M_{W2}$ contribution in the collinear phase (especially below 320 K) can be explained by a gradual re-ordering (re-orientation) of the conical longitudinal spin moments into the collinear ones near $T_{c2}$. Above 320 K, the corresponding coercive field $H_{C2}$ rapidly increases on heating and consequently, the $M_{W2}$ contribution decreases. Temperature dependence of $H_{C2}$ in the collinear phase can be well fitted to the critical law:

$$H_{C2} = H_0 \left( T/T_0 - 1 \right) \qquad (3)$$

with parameters $H_0 = 14$ kOe and $T_0 = T_{c2} = 305$ K (Fig. 2b). The $H_0$ value corresponds to the field (10 – 15 kOe) providing saturation of the integral magnetization at all temperatures.

## 3. Temperature evolution of the natural FMR

Frequency dependences of the transmission coefficient $Tr(f)$ of the coplanar waveguide (CPW) loaded with a CoZnU sample at temperatures above 200 K and in the frequency range near the $Tr(f)$ minimum corresponding to the natural FMR (the lowest frequency magnon mode marked as F4) are extracted from the data presented in [6] and are shown in Fig. 3. In the collinear phase, FMR splits into two components F4a and F4b. Other $Tr(f)$ minima correspond to the nearest magnetic modes which can interfere with the FMR. The F3 mode is attributed to dynamics of the inhomogeneous magnetic structure of the non-magnetized hexaferrite ceramics with various orientations of the local magnetic moments of grains and domains. The F5 and F6 modes are magnetodielectric with dominating influence of the magnetic properties on their temperature and field dependences (see [6] for details and reasons of the modes assignment).

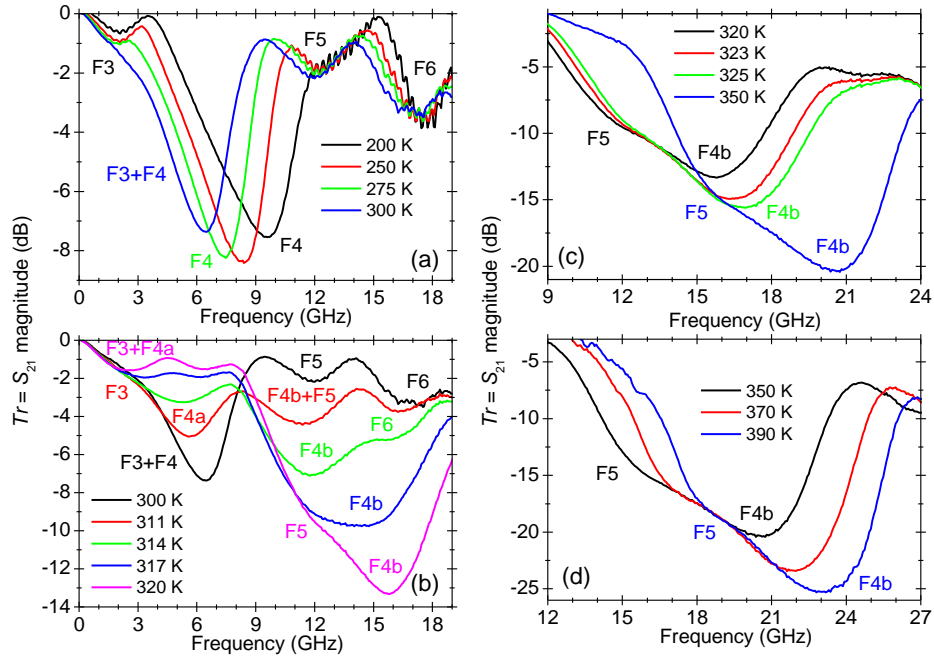

Fig. 3. Temperature evolution of the FMR and nearest magnetic modes: transmission coefficient $Tr$ spectra (a) in the conical phase below $T_{c2}$, (b) in the vicinity of $T_{c2}$, and (c, d) in the collinear phase above 320 K.

The F4 magnon in the conical phase below $T_{c2}$ and the F4b magnon in the collinear phase above 320 K clearly dominate the transmission spectra and are well separated from the other modes (Fig. 3a,c,d). Due to the low frequency (~6 GHz) at $T_{c2}$ and its quick increase on further heating, the F4b magnon interferes with the F5 and F6 modes in the temperature range between $T_{c2}$ and ~320 K (Fig. 3b). The interference is essential and visible due to the broad diffused $Tr(f)$ minima of all modes. The F6 mode is absent above 320 K in the transmission spectra, and the F5 is seen as lower-frequency shoulder of the dominating F4b component. The F3 mode is present both in the conical and collinear phases, mainly at frequencies ~ 2.5 GHz. Between $T_{c2}$ and 320 K, the F3 mode interferes with the F4b FMR component. Temperature dependences of the modes' frequencies are shown in Fig. 4.

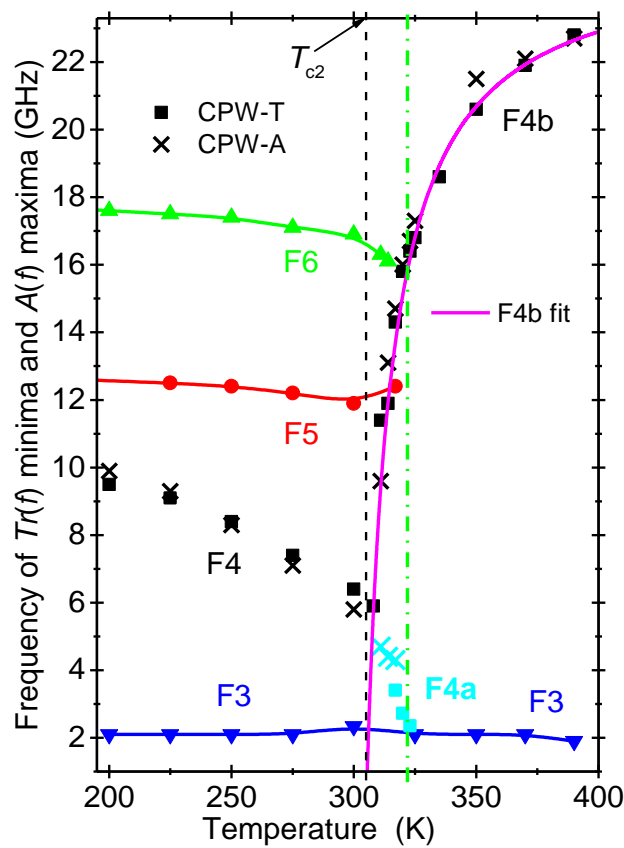

Fig. 4. Temperature evolution of the FMR (F4, F4a, F4b) and of the magnetic modes (F3, F5, F6) frequencies corresponding to the transmission coefficient minima and absorption coefficient maxima (filled symbols and crosses, respectively) of the CPW loaded with a CoZnU sample. A black dashed and a green dash-dot vertical lines correspond to $T_{c2}$ and 320 K, respectively. A magenta solid line corresponds to the F4b fit (Eq. 4).

The temperature range of the magnetic modes' interference (305 K – 320 K) corresponds well to the temperature range of a sharp (but not step-like) decrease of the weak-field ($H$ = 100 Oe) magnetization $M_W(T)$ on heating in the collinear phase above the $T_{c2}$ phase transition (Fig. 2a). This decrease is explained by the existence of the two $M_W$ contributions ($M_{W1}$ and $M_{W2}$) in the conical phase and only one $M_{W1}$ in the collinear phase.

All magnetic modes shown in Fig. 4 are excited by the weak electromagnetic field at zero magnetic bias field in the previously non-magnetized ceramics. Temperature evolution of their dynamics near the conical – collinear ferrimagnetic phase transition at $T_{c2}$ correlates with the evolution of the weak-field magnetization $M_W$ and its coercive field. The most sensitive and temperature-dependent is the FMR magnon. In the conical phase, the F4 frequency decreases down to ~6 GHz at $T_{c2}$ and then F4 splits in two components on further heating. The frequency of the F4a component decreases down to ~2.5 GHz at 320 K and joins the F3 mode. The frequency of the F4b component increases up to ~23 GHz at 390 K. Its critical temperature dependence is proportional with a constant factor $f_0 = 27.5$ GHz to that of the inverse magnetic permeability following the Curie-Wiess law:

$$f_{F4b} = {f_0}/{\mu'} = {f_0}/{(\mu_L + \frac{C}{T-T_0})}, \qquad (4)$$

with parameters $\mu_L = 1.085$, $C = 11$ K, and critical temperature $T_0 = T_{c2} = 305$ K which are close to the Curie-Wiess parameters for the 1 GHz permeability [5]. The $f_0$ value characterizes the $f_{F4b}$ at temperatures far above $T_{c2}$. The quickest $f_{F4b}$ increase (from 6 GHz to 16 GHz) is observed in the temperature range between $T_{c2}$ and 320 K where the $M_{W2}$ magnetization contribution gradually switches off.

Splitting of the FMR into the two components F4a and F4b in the collinear phase correlates with the splitting of the coercive fields of $H_{C1}$ and $H_{C2}$ (Fig. 2b). Both $f_{F4b}$ and $H_{C2}$ critically decrease towards $T_{c2}$. Generally, presence of the natural FMR evidences existence of the non-zero internal magnetization and magnetocrystalline anisotropy in the previously non-magnetized hexaferrite ceramics [6]. Both the FMR and coercive field splitting in the collinear phase are caused by a change of the magnetocrystalline anisotropy during conical to collinear phase transition at $T_{c2}$.

## 4. Correlation between the FMR and magnetization in the weak magnetic field

A weak magnetic bias field essentially influences the room temperature (RT) transmission $Tr$ and reflection loss $RL$ spectra of the CPW loaded with a CoZnU sample [6]. The most complicated and interesting behaviour shows the FMR mode F4 (Fig. 5). With increasing bias field $H$, the F4 transmission minimum first becomes deeper and less diffused, then above 280 Oe the minimum diffuses and its depth decreases, while the frequency of the minimum does not change essentially (Figs. 5a, 6). The diffused F4 *RL* minimum also first becomes deeper and sharper with increasing bias field *H*, then above 140 Oe it splits in two sharp minima (F4a and F4b). The further increase of *H* results in the depth redistribution of the *RL* split minima and finally to the diffusion and suppression of both of them (Figs 5b, 6).

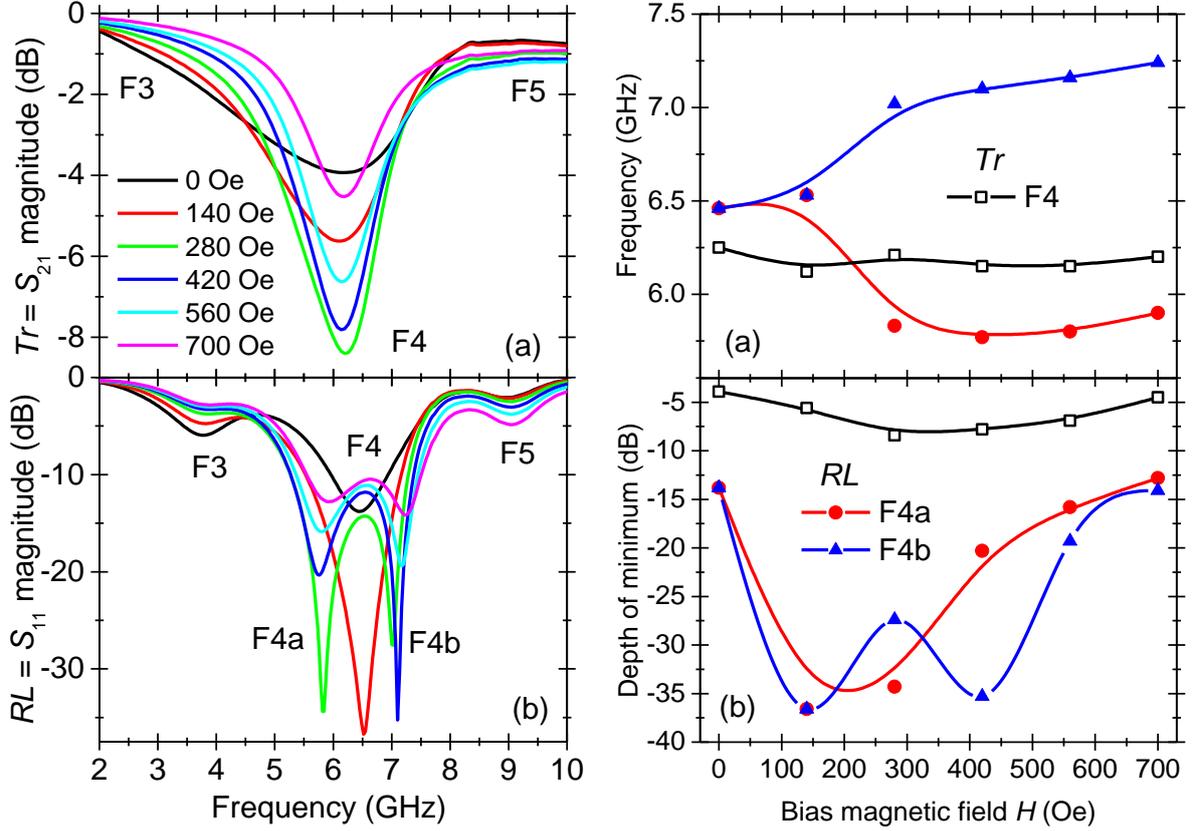

Fig. 5. Room-temperature transmission coefficient *Tr* spectra (a) and reflection loss *RL* spectra (b) of the CPW loaded with a CoZnU sample at different magnetic bias fields. The minima correspond to the FMR mode (F4, F4a and F4b) and to the modes which interfere with the FMR (F3 and F5).

Fig. 6. Room-temperature bias magnetic field dependences of the frequencies (a) and depth (b) of the *Tr* and *RL* minima corresponding to the F4, F4a and F4b modes in the CPW loaded with a CoZnU sample.

Change of the intensity, splitting and shift to higher frequencies of the MW magnetic excitations, especially FMR, by the applied bias field is usual and well known [2,3]. What is unusual and new, is that these phenomena are observed in the CoZnU hexaferrite under such low *H* values, below 700 Oe. It can be explained by the low coercive fields ($H_{C1}$ and $H_{C2}$) of the weak-field magnetization contributions in the conical phase at RT (Fig. 2b). Frequencies of both the *Tr*(*f*) and *RL*(*f*) minima characterize the magnetic modes' frequencies. Different presentation of the F4 mode in the transmission and reflection loss spectra is related to the different experimental setups [6]. The F4 *Tr*(*f*) minimum is less intense and more diffuse than the *RL*(*f*) one, and the *Tr*(*f*) minimum is defined by both the reflection and absorption, while *RL*(*f*) is defined only by the absorption. Therefore, splitting of the *Tr*(*f*) minimum in the low bias field is not observed. Nevertheless, the mean frequencies and widths of the *Tr*(*f*) minimum and the system of the F4a and F4b components are quite similar (Fig. 6). A non-monotonic change of the depth of F4a and F4b *RL* minima with increasing bias field reflects redistribution of the electromagnetic absorption between two FMR components.

Splitting of the FMR magnon under the application of weak bias magnetic field (Figs. 5, 6) looks similar to the natural FMR splitting caused by the conical to collinear phase transition (Fig. 4). Nevertheless, even considering that $T_{c2}$ is close to RT, we cannot explain the *RL* minimum splitting by the bias-induced phase transition. The applied bias is too weak. According to the magnetization loops [5], the field-induced conical to collinear phase transitions is observed at much higher fields, at least above 2 kOe. We attribute the observed splitting to the presence of two weak-field magnetization contributions with low coercive fields in the conical phase. The applied bias is high enough to split the weak-magnetization components and, consequently, to split the FMR.

The *Tr* and *RL* magnitudes are used for the characterization of the microwave shielding efficiency (*SE*) and absorption (*MA*) of materials [10,11,12]: *SE* = - *Tr* (dB) and *MA* = -*RL* (dB). The FMR mode contributes to both parameters at the 5-7 GHz frequency range due to the strong interaction with electromagnetic waves. At RT and zero bias field, both these contributions are relatively small: *SE* = 4 dB, *MA* = 15 dB, but application of the weak magnetic field increases *SE* up to 8 dB and *MA* up to 37 dB (Fig. 5). Also heating results in the *SE* increase up to 13 dB at 320 K and 25 dB at 390 K (Fig. 3). Frequency of the *SE* maximum shifts to 24 GHz and broadens. Taking into account that the 20 dB level of the *SE* and *MA* means the 10 times improvement of the shielding and absorption, we consider the CoZnU hexaferrite as a suitable MW absorbing material.

**Conclusions**

Temperature dependence of the FMR near $T_{c2}$ without application of the bias magnetic field and its change under application of the weak bias (*H* < 700 Oe) at room temperature (i.e., in the conical phase close to $T_{c2}$) were analysed in detail. The FMR behaviour and interference with other MW magnons at temperatures 305 – 320 K is found to correlate with the temperature evolution of the weak-field magnetization and its coercive field: from two components in the conical phase to one component in the collinear phase. The observed FMR splitting above $T_{c2}$ correlates with the splitting of the coercive fields of two components of the weak-field magnetization. A high sensitivity of the FMR dynamics to the weak magnetic bias field in the conical phase close to $T_{c2}$ is shown to be caused by the gradual transformation of the mixed transverse and longitudinal conical magnetic moments to the predominantly longitudinal ones with increasing the bias field. Application of the weak magnetic bias field allows to adjust the MW absorption, and its level of above 30 dB is achieved in the coplanar waveguide near the FMR frequency (5.5 – 7.5 GHz), that allows to consider the $Sr_4CoZnFe_{36}O_{60}$ hexaferrite ceramics as a possible MW absorbing material working near the room temperature.

**Acknowledgments**

This work has been supported by the Czech Science Foundation (Project No. 21-06802S), Grant Agency of the Czech Technical University in Prague (Project No. SGS22/182/OHK4/3T/14), by the Research Infrastructure NanoEnviCz (funded by MEYS CR, Projects No. LM2018124), and by project TERAFIT - CZ.02.01.01/00/22_008/0004594 co-financed by European Union and the Czech Ministry of Education, Youth and Sports.

**Declaration of Competing Interest**

The authors report there are no competing interests to declare.

**Data availability**

Data will be made available upon reasonable request.